\documentclass[12pt,draftcls,onecolumn]{IEEEtran} 
\usepackage{amsmath,graphicx}
\usepackage{amsfonts}
\usepackage{enumerate}
\usepackage{algorithm} 
\usepackage{algorithmic} 

\usepackage{mathrsfs}
\usepackage{stfloats}
\usepackage{enumerate}
\usepackage{epstopdf}
\usepackage{subfigure}
\usepackage{float}
\usepackage{amsbsy}

\ifCLASSINFOpdf

\else

\fi

\begin{document}

\title{Simultaneous Transmission and Reflection Reconfigurable Intelligent Surface Assisted MIMO Systems}

\author{Hehao Niu, Zheng Chu, \emph{Member, IEEE}, Fuhui Zhou, \emph{Senior Member, IEEE}, \\
           Pei Xiao, \emph{Senior Member, IEEE}, and Naofal Al-Dhahir, \emph{Fellow, IEEE}

        \thanks{
%

         H. Niu is with the Institute of Electronic Countermeasure, National University of Defense Technology, Hefei, 230037, China, (e-mail: niuhaonupt@foxmail.com).

Z. Chu, and P. Xiao are with the Institute for Communication Systems, University of Surrey, Guildford GU2 7XH, U.K. (e-mail: andrew.chuzheng7@gmail.com, p.xiao@surrey.ac.uk).

   F. Zhou is with the College of Electronic and Information Engineering, Nanjing University of Aeronautics and Astronautics, Nanjing, 210000, China (email: zhoufuhui@ieee.org).

N. Al-Dhahir is with the Department of Electrical and Computer Engineering, The University of Texas at Dallas, Richardson, TX 75080 USA. (email:aldhahir@utdallas.edu). }

}

\markboth{IEEE xxx,~Vol.~xx, No.~xx, ~2021}%
{Niu \MakeLowercase{\textit{et al.}}: }
\maketitle

\begin{abstract}
In this work, we investigate a novel simultaneous transmission and reflection reconfigurable intelligent surface (RIS)-assisted multiple-input multiple-output downlink system, where three practical transmission protocols, namely, energy splitting (ES), mode selection (MS), and time splitting (TS), are studied. For the system under consideration, we maximize the weighted sum rate with multiple coupled variables. To solve this optimization problem, a block coordinate descent algorithm is proposed to reformulate this problem and design the precoding matrices and the transmitting and reflecting coefficients (TARCs) in an alternate manner. Specifically, for the ES scheme, the precoding matrices are solved using the Lagrange dual method, while the TARCs are obtained using the penalty concave-convex method. Additionally, the proposed method is extended to the MS scheme by solving a mixed-integer problem. Moreover, we solve the formulated problem for the TS scheme using a one-dimensional search and the Majorization-Minimization technique. Our simulation results reveal that: 1) Simultaneous transmission and reflection RIS (STAR-RIS) can achieve better performance than reflecting-only RIS; 2) In unicast communication, TS scheme outperforms the ES and MS schemes, while in broadcast communication, ES scheme outperforms the TS and MS schemes.
\end{abstract}


\IEEEpeerreviewmaketitle
\setlength{\baselineskip}{1\baselineskip}
\newtheorem{fact}{Fact}
\newtheorem{assumption}{Assumption}
\newtheorem{lemma}{\underline{Lemma}}[section]
\newtheorem{corollary}{\underline{Corollary}}[section]
\newtheorem{proposition}{\underline{Proposition}}[section]
\newtheorem{example}{\underline{Example}}[section]
\newtheorem{remark}{\underline{Remark}}[section]
\newcommand{\mv}[1]{\mbox{\boldmath{$ #1 $}}}

\section{Introduction}
Recently, the reconfigurable intelligent surface (RIS) has attracted great interest from both industry and academia. Since RIS can be installed on flat surfaces to reflect wireless signals and establish a virtual end-to-end link between the transmitter (Tx) and the receiver (Rx), RIS has emerged as a key technology to overcome blockage in wireless systems \cite{Liu2020}. RIS has sparked significant interest in various wireless networks and applications such as the multiple-input single-output (MISO) channel in \cite{MuTC2021}, multi-cell networks in \cite{NiTWC2021}, multiple-input multiple-output (MIMO) networks in \cite{ZhangTVT20201}, unmanned aerial vehicle (UAV) networks in \cite{Mu20211}, and the non-orthogonal multiple access (NOMA) channel in \cite{ZuoTC2020}, among others.

However, most existing works consider the reflecting-only RIS, where the RIS can only reflect the incident wireless signal. As such, Tx and Rx need to be deployed on the same side of the RIS, which may not be practical \cite{Zhang2021}. Fortunately, with the development of the reflective-transmissive meta-surfaces, a promising technique called simultaneous transmission and reflection RIS (STAR-RIS) was proposed, which can split the incident signal into two parts \cite{ZhangTVT2020}. To be specific, the signal transmitted to the region behind the RIS is referred to as the transmitted signal, while the signal reflected to the region in front of the RIS is referred to as the reflected signal \cite{Zeng2020}. By altering the electromagnetic properties of the STAR-RIS elements, two groups of independent coefficients, termed as the transmitting and reflecting coefficients (TARCs), can be designed to control the transmitted and reflected signals \cite{Xu2021}. It was shown in \cite{Liu2021} that the STAR-RIS can extend the coverage of a wireless network. In \cite{Mu2021}, three protocols were proposed to coordinate the transmission (T) mode and the reflection (R) mode in STAR-RIS-aided MISO networks, with the aim to minimize the transmit power. However, the synergy between the T and R modes for the STAR-RIS-aided MIMO network has not been studied yet.

Against this background, this work investigates joint precoding matrices and TARCs optimization in STAR-RIS-aided MIMO networks. Firstly, we utilize an equivalent weighted minimum mean-squared error (WMMSE) method to reformulate the weighted sum rate (WSR) objective. Then, for the energy splitting (ES) scheme, the precoding matrices are found by the Lagrange dual method, while the TARCs are obtained by the penalty concave-convex method. Moreover, the proposed algorithm is extended to the mode selection (MS) scheme by solving a mixed-integer optimization problem. Furthermore, for the time splitting (TS) scheme, the formulated problem is solved by applying a one-dimensional search and the Majorization-Minimization (MM) approach. Finally, our simulation results demonstrate the superiority of the proposed approach and reveal that: 1) STAR-RIS outperforms the reflecting-only RIS; 2) In unicast communication, the TS scheme outperforms the ES and MS schemes, while in broadcast communication, the ES scheme outperforms the TS and MS schemes.

The main difference of our work with the most related work \cite{Mu2021} are summarized as follows:
1) Firstly, in terms of the system model and problem formulation, our work investigate the STAR-RIS-assisted MIMO network with the aim to maximize the WSR. While \cite{Mu2021} handled the power minimization design in a STAR-RIS-aided MISO network; 2) From the point of the optimization method, our method mainly based on the Lagrange dual, penalty concave-convex method, and the alternating optimization (AO). While \cite{Mu2021} mainly based on the successive inner-approximation and semidefinite relaxation (SDR).

The rest of this work is organized as follows: Section \ref{secSM} introduces the signal model and three protocols of the STAR-RIS. Section \ref{SMPF} provides the system model and problem formulations. An efficient algorithm is proposed for the ES scheme in Section \ref{Design}, and extended to the MS and TS schemes in Section \ref{Ext}. Section \ref{SiRe} shows the simulation results and conclusion is given in Section \ref{Con}.
\section{Signal Model}\label{secSM}
This section describes the signal model of STAR-RIS and its three T and R protocols.

\subsection{Signal Model of STAR-RIS}
Let $M$ be the number of the STAR-RIS elements. The transmitted and reflected signals by the $m$-th element, $m \in {\cal M} \buildrel \Delta \over = \left\{ {1,2, \ldots ,M} \right\}$, are given as ${t_m} = \sqrt {\alpha _m^t} {e^{j\phi _m^t}}{s_m}$, and ${r_m} = \sqrt {\alpha _m^r} {e^{j\phi _m^r}}{s_m}$, respectively, where $s_m$ represents the signal incident on the $m$-th element, $\sqrt {\alpha _m^t}  \in \left[ {0,1} \right],\phi _m^t \in \left[ {0,2\pi } \right)$ and $\sqrt {\alpha _m^r}  \in \left[ {0,1} \right],\phi _m^r \in \left[ {0,2\pi } \right)$ are the amplitude and phase shift responses of the $m$-th element, respectively. It should be noted that, for each element, $\phi _m^t$ and $\phi _m^r$ can be chosen independently. However, $\sqrt {\alpha_m^t} $ and $\sqrt {\alpha _m^r} $ satisfy the energy relationship $\alpha _m^t + \alpha _m^r = 1,\forall m \in {\cal M}$, \cite{Mu2021}.

\subsection{Three Protocols for STAR-RIS}
Here, we briefly describe the three existing operating schemes for STAR-RIS. Readers can refer to \cite{Mu2021} for more details:

1) In the ES protocol, all elements simultaneously operate in the two modes. Thus, the TARCs are modelled as ${\bf{\Phi }} _t^{ES} = {\rm{Diag}}\left( {\sqrt {\alpha _1^t} {e^{j\phi _1^t}}, \ldots ,\sqrt {\alpha _M^t} {e^{j\phi _M^t}}} \right)$, and ${\bf{\Phi }} _r^{ES} = {\rm{Diag}}\left( {\sqrt {\alpha _1^r} {e^{j\phi _1^r}}, \ldots ,\sqrt {\alpha _M^r} {e^{j\phi _M^r}}} \right)$, where $\alpha _m^t,\alpha _m^r \in \left[ {0,1} \right]$, $\alpha _m^t + \alpha _m^r = 1$, and $\phi _m^t,\phi _m^r \in \left[ {0,2\pi } \right), \forall m \in {\cal M}$. For the ES scheme, since the TARCs of each element can be optimized, a high degree of system flexibility is enabled. However, the increased variables lead to a higher overhead for information exchange between Tx and the RIS.

2) In the MS protocol, all elements are separated into two parts, namely, one part with $M_t$ elements working in the T mode, while the other part with $M_r$ elements working in the R mode, where ${M_t} + {M_r} = M$. Thus, the TARCs are given by ${\bf{\Phi }} _t^{MS} = {\rm{Diag}}\left( {\sqrt {\alpha _1^t} {e^{j\phi _1^t}}, \ldots ,\sqrt {\alpha _M^t} {e^{j\phi _M^t}}} \right)$, and ${\bf{\Phi }} _r^{MS} = {\rm{Diag}}\left( {\sqrt {\alpha _1^r} {e^{j\phi _1^r}}, \ldots ,\sqrt {\alpha _M^r} {e^{j\phi _M^r}}} \right)$, respectively, where $\alpha _m^t,\alpha  _m^r \in \left\{ {0,1} \right\},\alpha _m^t + \alpha _m^r = 1$, and $\phi _m^t,\phi _m^r \in \left[ {0,2\pi } \right),\forall m \in {\cal M}$. In fact, MS can be treated as a special case of ES, since the amplitude coefficients for the T and R modes are restricted to binary values, which may cause certain performance loss. However, such ``on-off" operation is easier to implement when compared to the ES protocol.

3) In the TS protocol, the STAR-RIS switches all elements between the two modes in different orthogonal time durations. Let $ {\tau_t} $ and ${\tau_r}$ be the time allocation to the T and R modes, respectively, satisfying ${\tau_t} + {\tau_r} = 1$. Thus, the TARCs are given by ${\bf{\Phi }} _t^{TS} = {\rm{Diag}}\left( {{e^{j\phi _1^t}},{e^{j\phi _2^t}}, \ldots ,{e^{j\phi _M^t}}} \right)$ and ${\bf{\Phi }} _r^{TS} = {\rm{Diag}}\left( {{e^{j\phi _1^r}},{e^{j\phi  _2^r}}, \ldots ,{e^{j\phi  _M^r}}} \right)$, respectively, where $\phi  _m^t,\phi  _m^r \in \left[ {0,2\pi } \right),\forall m \in {\cal M} $.
The design for the TS scheme is simpler, since the TARCs are not coupled. However, the switching between the two modes requires precise synchronization, which incurs a higher implementation complexity.

\section{System Model and Problem Formulation} \label{SMPF}
In this section, a STAR-RIS assisted MIMO system is considered first and then we formulate the WSR objective for the three protocols.

\subsection{System Model}
A STAR-RIS assisted MIMO system is investigated, where a Tx intends to establish a link with two users with the assistance of a STAR-RIS equipped with $M$ elements. We assume that both regions, i.e., the T and R regions, are simultaneously served by the RIS due to its hardware model characteristics \cite{Mu2021}. Particularly, one user located in the T region is referred to as the T user, while the other user in the R region is known as the R user. The direct links between Tx and the users are assumed negligible due to blockage. Let ${\bf{F}} \in {\mathbb{C}^{M \times N}}$, ${\bf{H}}_t \in {\mathbb{C}^{ {N_t} \times M}}$, and ${\bf{H}}_r \in {\mathbb{C}^{{N_r} \times M}}$ denote the channel coefficients matrices between Tx and STAR-RIS, STAR-RIS and the T user, STAR-RIS and the R user, respectively, where $N$, $N_t$ and $N_r$ are the numbers of antennas for Tx, the T user and the R user, respectively. In addition, perfect channel state information (CSI) of these links is assumed to be available at Tx since we aim to derive a performance upper bound for the network.

Let ${{\bf{s}}_l} \in {\mathbb{C}^{{N_{{d_l}}} \times 1}}, l \in \left\{ {t,r} \right\}$ be the transmit symbols for the $l$-th user with ${N_{{d_l}}}$ being the corresponding number of data streams. Then, for the ES and MS schemes, the signal received by the $l$-th user is given as
\begin{equation}\label{eq:recsig}
{{\bf{y}}_l} = {{\bf{H}}_l}{{\bf{\Phi }}_l}{\bf{F}}\sum\nolimits_l {{{\bf{W}}_l}{{\bf{s}}_l}}  + {{\bf{n}}_l},
\end{equation}
where ${{\bf{W}}_l} \in {\mathbb{C}^{N \times {{N_{{d_l}}}}}}$ denotes the precoding matrix for the $l$-th user, ${\bf{\Phi }}_l \in {\mathbb{C}^{M \times M}}$ is the STAR-RIS coefficient matrix, ${{\bf{n}}_l} \in {\mathbb{C}^{{N_l} \times 1}} $ is the noise at the $l$-th user with \({\bf{n}}_l \sim {\cal C}{\cal N}\left( {{\bf{0}}, \sigma_l^2{\bf{I}} } \right)\), and $\sigma_l^2$ is the noise power. Here, we assume that ${{\bf{s}}_t}$ and ${{\bf{s}}_r}$ are independent, since we study unicast communication. On the other hand, for broadcast communication, the Tx sends the same information to the users. In the following, we mainly focus on the unicast scenario. However, the proposed method can be extended the broadcast scenario.

By denoting ${{{{\bf{\bar H}}}_l} = {{\bf{H}}_l}{{\bf{\Phi}}_l}{\bf{F}}}$, the achievable information rate for the $l$-th user is given by
\begin{equation}\label{eq:rate}
{R_l} = {\log_2} \left| {{\bf{I}} + {{{\bf{\bar H}}}_l}{{\bf{W}}_l}{\bf{W}}_l^H{\bf{\bar H}}_l^H{\bf{C}}_l^{ - 1}} \right|,
\end{equation}
where ${{\bf{C}}_l} = {{{\bf{\bar H}}}_l}{{\bf{W}}_{{l^ \prime }}}{\bf{W}}_{{l^ \prime }}^H{\bf{\bar H}}_l^H + \sigma _l^2{\bf{I}}$ is the interference plus noise covariance matrix. When $l=r, l^ \prime=t$, and vice versa.

On the other hand, for the TS scheme, since different users are allocated different time slots, the information rate for the $l$ user is given by
\begin{equation}\label{eq:rateTs}
{R_l} = {\tau _l}{\log _2}\left| {{\bf{I}} + \sigma _l^{ - 2}{{{\bf{\bar H}}}_l}{{\bf{W}}_l}{\bf{W}}_l^H{\bf{\bar H}}_l^H} \right|.
\end{equation}

\subsection{Optimization Problem Formulation}
Our goal is to maximize the WSR by jointly designing the precoding matrices and the TARCs. Specifically, for the ES scheme, the problem is formulated as:
\begin{subequations}\label{eq:Op}
\begin{align}
\mathop {\rm{ES:}}\;\;&\mathop {\max }\limits_{{{\bf{W}}_l},{{\bf{\Phi}}_l}} \;\;\sum\nolimits_l {{w_l}{R_l}}  \label{eq:Opo} \\
&\;\;\;\;{\rm{s.t.}}\;\; \sum\nolimits_l {{\rm{Tr}}\left( {{{\bf{W}}_l}{\bf{W}}_l^H} \right)}  \le {P_s},\label{eq:Opc1} \\
&~{\left[ {{{\bf{\Phi }}_l}} \right]_m} = \sqrt {\alpha _m^l} {e^{j\phi _m^l}},\alpha _m^l \in \left[ {0,1} \right],\phi _m^l \in \left[ {0,2\pi } \right), \label{eq:Opc2}
\end{align}
\end{subequations}
where ${{w_l} \in \left[ {0,1} \right],\sum\nolimits_l {{w_l} = 1} }$ is the weight for the $l$-th user, and $P_s$ denotes the transmit power budget. Note that \eqref{eq:Op} can be reformulated to the MS scheme by replacing $\alpha _m^l \in \left[ {0,1} \right]$ with $\alpha _m^l \in \left\{ {0,1} \right\}$.

On the other hand, for the TS scheme, we formulate the following WSR problem
\begin{subequations}\label{eq:TsOp}
\begin{align}
\mathop {\rm{TS:}}&\;\;\mathop {\max }\limits_{{{\bf{W}}_l},{{\bf{\Phi }}_l},{\tau _l}} \;\;\sum\nolimits_l {{\tau_l}{w_l}{R_l}}\label{eq:TsOpo} \\
&\;\;\;\;\;\;{\rm{s.t.}}\;\;\sum\nolimits_l {{\tau _l}{\rm{Tr}}\left( {{{\bf{W}}_l}{\bf{W}}_l^H} \right)}\le {P_s},\label{eq:TsOpc1} \\
&~~~~~~~~~~~{\left[ {{{\bf{\Phi }}_l}} \right]_m}= {e^{j\phi _m^l}},\phi _m^l \in \left[ {0,2\pi } \right), \label{eq:TsOpc2}  \\
&~~~~~~~~~~~{\tau ^l} \in \left[ {0,1} \right],\sum\nolimits_l {{\tau ^l}}  = 1. \label{eq:TsOpc3}
\end{align}
\end{subequations}

\section{Joint Precoding and TARCs Design for The ES Scheme} \label{Design}
In this section, we investigate joint precoding and TARCs design for the ES scheme, where the formulated method is applicable to the design of the MS and TS schemes.

\subsection{Reformulation of Problem \eqref{eq:Op}}
In this subsection, we transform the intractable objective function (OF) \eqref{eq:Opo} into an equivalent form. Then, we propose a block coordinate descent (BCD)-based method to solve it.

Firstly, we use a linear decoding matrix ${\bf{U}}_l \in {\mathbb{C}^{{N_l} \times N}}$ to recover the signal vector ${{{\bf{\hat s}}}_l}$ for the $l$-th user ${{{\bf{\hat s}}}_l} = {\bf{U}}_l^H{{\bf{y}}_l}$. The mean-squared error matrix of the $l$-th user is thus given by
\begin{equation}\label{eq:MSE}
\begin{split}
&{{{\bf{E}}_l} = \mathbb{E}\left[ {\left( {{{{\bf{\hat s}}}_l} - {{\bf{s}}_l}} \right){{\left( {{{{\bf{\hat s}}}_l} - {{\bf{s}}_l}} \right)}^H}} \right]}\\
& ~~~= \left( {{\bf{U}}_l^H{{{\bf{\bar H}}}_l}{{\bf{W}}_l} - {\bf{I}}} \right){{\left( {{\bf{U}}_l^H{{{\bf{\bar H}}}_l}{{\bf{W}}_l} - {\bf{I}}} \right)}^H}\\
& ~~~+ {\bf{U}}_l^H{{{\bf{\bar H}}}_l}{{\bf{W}}_l}{\bf{W}}_l^R{\bf{\bar H}}_l^H{{\bf{U}}_l} + \sigma _l^2{\bf{U}}_l^H{{\bf{U}}_l}.
\end{split}
\end{equation}

Upon introducing a set of auxiliary matrices ${\bf{V}}_l \in {\mathbb{C}^{N \times N}}$, problem \eqref{eq:Op} can be reformulated as follows
\begin{subequations}\label{eq:Fp}
\begin{align}
&\;\;\mathop {\max }\limits_{{{\bf{W}}_l},{{\bf{\Phi }}_l}} \;\;\sum\nolimits_l {{w_l}{h_l}\left( {{{\bf{V}}_l},{{\bf{U}}_l},{{\bf{W}}_l},{{\bf{\Phi }}_l}} \right)} \label{eq:Fpo} \\
&\;\;\;\;{\rm{s.t.}}\;\; \eqref{eq:Opc1},\eqref{eq:Opc2}, \label{eq:Fpc1}
\end{align}
\end{subequations}
where ${h_l\left( {{{\bf{V}}_l},{{\bf{U}}_l},{{\bf{W}}_l},{{\bf{\Phi }}_l}} \right) = {\log_2} \left| {{{\bf{V}}_l}} \right| - {\rm{Tr}}\left( {{{\bf{V}}_l}{{\bf{E}}_l}} \right) + {d_l}}$.

With fixed $\left\{ {{{\bf{W}}_l},{{\bf{\Phi }}_l}} \right\}$, the optimal matrix variables $\left\{ {{{\bf{U}}_l},{{\bf{V }}_l}} \right\}$ are given by
\begin{subequations}\label{eq:VU}
\begin{align}
&{{\bf{U}}_l^ \star  = {{\left( {{{{\bf{\bar H}}}_l}{{\bf{W}}_l}{\bf{W}}_l^H{\bf{\bar H}}_l^H + {{\bf{C}}_l}} \right)}^{ - 1}}{{{\bf{\bar H}}}_l}{{\bf{W}}_l}}, \label{eq:U} \\
&{\bf{V}}_l^ \star  = {\bf{E}}_l^{ \star  - 1}, \label{eq:V}
\end{align}
\end{subequations}
where ${\bf{E}}_l^ \star$ is obtained by substituting ${\bf{U}}_l^ \star$ into \eqref{eq:MSE}, i.e.,
${{\bf{E}}_l^ \star  = {{\bf{I}}_l} - {\bf{W}}_l^H{\bf{\bar H}}_l^H{{\left( {{{{\bf{\bar H}}}_l}{{\bf{W}}_l}{\bf{W}}_l^H{\bf{\bar H}}_l^H + {{\bf{C}}_l}} \right)}^{ - 1}}{{{\bf{\bar H}}}_l}{{\bf{W}}_l}}$ \cite{ZhangTVT2020}.

Next, we optimize $\left\{ {{{\bf{W}}_l},{{\bf{\Phi }}_l}} \right\}$ for the given $\left\{ {{{\bf{U}}_l},{{\bf{V}}_l}} \right\}$, where the BCD method is used to alternately optimize $\left\{ {{{\bf{W}}_l},{{\bf{\Phi }}_l}} \right\}$ efficiently.

\subsection{Optimization of The Precoding Matrices}
Here, we fix $\left\{ {{{\bf{U}}_l},{{\bf{V}}_l},{{\bf{\Phi }}_l}} \right\}$ and focus on the subproblem with respect to (w.r.t.) ${\bf{W}}_l$. Specifically, by substituting ${\bf{E}}_l^ \star$ into \eqref{eq:Fpo} and neglecting the terms which are not related to ${\bf{W}}_l$, the following problem can be formulated \cite{ZhangTVT20201}
\begin{subequations}\label{eq:WrWt}
\begin{align}
&\mathop {\min }\limits_{{{\bf{W}}_l}} \;\;\sum\nolimits_l {{\rm{Tr}}\left( {{\bf{W}}_l^H{\bf{A}}{{\bf{W}}_l}} \right) - 2\Re \left\{ {{\rm{Tr}}\left( {{{\bf{B}}_l}{{\bf{W}}_l}} \right)} \right\}} \label{eq:WrWto} \\
&\;\;{\rm{s.t.}}\;\;\eqref{eq:Opc1},  \label{eq:WrWtc1}
\end{align}
\end{subequations}
where ${\bf{A}} = \sum\nolimits_l {{w_l}{\bf{\bar H}}_l^H{{\bf{U}}_l}{{\bf{V}}_l}{\bf{U}}_l^H{{{\bf{\bar H}}}_l}} $, and ${{\bf{B}}_l} = {w_l}{{\bf{V}}_l}{\bf{U}}_l^H{{\bf{\bar H}}_l}$, respectively. Moreover, the Lagrange cost function of \eqref{eq:WrWt} is
\begin{equation}\label{eq:lag}
\begin{split}
{\cal L}\left( {{{\bf{W}}_l},\lambda } \right) &= \sum\nolimits_l {{\rm{Tr}}\left( {{\bf{W}}_l^H{\bf{A}}{{\bf{W}}_l}} \right) - 2\Re \left\{ {{\rm{Tr}}\left( {{{\bf{B}}_l}{{\bf{W}}_l}} \right)} \right\}}  \\
&+ \lambda \left( {\sum\nolimits_l {{\rm{Tr}}\left( {{{\bf{W}}_l}{\bf{W}}_l^H} \right)}  - {P_s}} \right).
\end{split}
\end{equation}
By the first order optimization condition, we have ${{\bf{W}}_l} = {\left( {{\bf{A}} + \lambda {\bf{I}}} \right)^{ - 1}}{\bf{B}}_l^H$. While the dual variable $\lambda$ can be found by the bisection search method in \cite{ZhangTVT20201}; we omit the details for brevity.


\subsection{Optimization of The TARCs}
Here, we focus on optimizing ${{\bf{\Phi }}_l}$ for given ${{\bf{W }}_l}$. Using a similar method as in \cite{ZhangTVT20201}, we formulate the following problem
\begin{subequations}\label{eq:PsRC}
\begin{align}
&\mathop {\min }\limits_{{{\boldsymbol{\phi }}_l}} \sum\nolimits_l {{\boldsymbol{\phi }}_l^H{{\bf{Z}}_l}{{\boldsymbol{\phi }}_l} - 2\Re \left\{ {{\boldsymbol{\phi }}_l^H{\bf{z}}_l^ * } \right\}} \label{eq:PsRCo}  \\
&\;\;{\rm{s.t.}}\;\;{\rm{diag}}\left( {{\boldsymbol{\phi }}_r^H{{\boldsymbol{\phi }}_r} + {\boldsymbol{\phi }}_t^H{{\boldsymbol{\phi}}_t}} \right) = {\bf{1}}, \label{eq:PsRCc1}
\end{align}
\end{subequations}
where ${\boldsymbol{\phi }}_l = {\rm{diag}}\left( {{{\bf{\Phi }} _l}} \right)$, and
\begin{subequations}\label{eq:Zlzl}
\begin{align}
&{{\bf{Z}}_l} = \left( {{w_l}{\bf{H}}_l^H{{\bf{U}}_l}{{\bf{V}}_l}{\bf{U}}_l^H{{\bf{H}}_l}} \right) \odot {\left( {{\bf{F}}\sum\nolimits_l {{{\bf{W}}_l}{\bf{W}}_l^H} {{\bf{F}}^H}} \right)^T}, \label{eq:Zl} \\
&{{\bf{z}}_l} = {\rm{diag}}\left( {{w_l}{\bf{F}}{{\bf{W}}_l}{{\bf{V}}_l}{\bf{U}}_l^H{{\bf{H}}_l}}\right). \label{eq:zl}
\end{align}
\end{subequations}

The most challenging condition is \eqref{eq:PsRCc1}, which is non-convex. To address this challenge, we introduce the matrices ${{\bf{\Omega }}_l} ,\forall l \in \left\{ {r,t} \right\}$, and obtain the following problem
\begin{subequations}\label{eq:RePsRC}
\begin{align}
&\mathop {\min }\limits_{{{\boldsymbol{\phi }}_l},{{\bf{\Omega }}_l}} \sum\nolimits_l {{\boldsymbol{\phi }}_l^H{{\bf{Z}}_l}{{\boldsymbol{\phi }}_l} - 2\Re \left\{ {{\boldsymbol{\phi }}_l^H{\bf{z}}_l^ * } \right\}} \label{eq:RePsRCo} \\
&\;\;{\rm{s.t.}}\;\;{\boldsymbol{\phi }}_r^H{{\boldsymbol{\phi}}_r} = {{\bf{\Omega }}_r},{\boldsymbol{\phi }}_t^H{{\boldsymbol{\phi }}_t} = {{\bf{\Omega }}_t}, \label{eq:RePsRCc1} \\
&\;\;\;\;\;\;\;\;\;\;{\rm{diag}}\left( {{{\bf{\Omega }}_r} + {{\bf{\Omega }}_t}} \right) = {\bf{1}}. \label{eq:RePsRCoc2}
\end{align}
\end{subequations}

However, \eqref{eq:RePsRCc1} is still non-convex. The following Lemma is useful to handle \eqref{eq:RePsRCc1}.

{\textit{Lemma 1 \cite{RashidTC2014}:}} The equality \({\bf{T}} = {\bf{t}}{{\bf{t}}^H}\) is equivalent to
\begin{equation}
\left\{ {\begin{array}{*{20}{c}}
{\left[ {\begin{array}{*{20}{c}}
{{{\bf{D}}_1}}&{\bf{T}}&{\bf{t}}\\
{{{\bf{T}}^H}}&{{{\bf{D}}_2}}&{\bf{t}}\\
{{{\bf{t}}^H}}&{{{\bf{t}}^H}}&1
\end{array}} \right]\succeq{\bf{0}}}\\
{{\rm{Tr}}\left( {{\bf{t}}{{\bf{t}}^H} - {{\bf{D}}_1}} \right) \ge 0},
\end{array}} \right.
\end{equation}
where \({{\bf{D}}_1} \in\mathbb{H} {^{N \times N}}\) and \({{\bf{D}}_2} \in\mathbb{H} {^{N \times N}}\) are slack variables.

With Lemma 1, \eqref{eq:RePsRC} can be recast as
\begin{subequations}\label{eq:FePsRC}
\begin{align}
&\mathop {\min }\limits_{{{\boldsymbol{\phi }}_l},{{\bf{\Omega }}_l}} \sum\nolimits_l {{\boldsymbol{\phi }}_l^H{{\bf{Z}}_l}{{\boldsymbol{\phi}}_l} - 2\Re \left\{ {{\boldsymbol{\phi }}_l^H{\bf{z}}_l^ * } \right\}} \label{eq:FePsRCo} \\
&\;\;{\rm{s.t.}}\;\;\;{\rm{diag}}\left( {{{\bf{\Omega }}_r} + {{\bf{\Omega }}_t}} \right) = {\bf{1}}, \label{eq:FePsRCc1}\\
&\;\;\;\;\;\;\;\;\;\;\;{\left[ {\begin{array}{*{20}{c}}
{{{\bf{D}}_{1,l}}}&{{{\bf{\Omega }}_l}}&{{{\boldsymbol{\phi }}_l}}\\
{{\bf{\Omega }}_l^H}&{{{\bf{D}}_{2,l}}}&{{{\boldsymbol{\phi}}_l}}\\
{{\boldsymbol{\phi }}_l^H}&{{\boldsymbol{\phi }}_l^H}&1
\end{array}} \right]\succeq{\bf{0}}},\label{eq:FePsRCc2} \\
&\;\;\;\;\;\;\;\;\;\;{{\rm{Tr}}\left( {{{\bf{D}}_{1,l}} - {{\boldsymbol{\phi }}_l}{\boldsymbol{\phi }}_l^H} \right) \le 0}. \label{eq:FePsRCc3}
\end{align}
\end{subequations}

However, \eqref{eq:FePsRCc3} is non-convex, but can be approximated as ${\rm{Tr}}\left( {{{\bf{D }}_{1,l}}} \right) \le 2{\rm{Tr}}\left( {{\boldsymbol{\phi }}_l^{\left( n \right)}{\boldsymbol{\phi}}_l^H} \right) - {\rm{Tr}}\left( {{\boldsymbol{\phi }}_l^{\left( n \right)}{{\left( {{\boldsymbol{\phi }}_l^{\left( n \right)}} \right)}^H}} \right)$ around given point ${\boldsymbol{\phi}}_l^{\left( n \right)}$, where $n$ denotes the number of iterations. Thus, we formulate the following convex problem
\begin{subequations}\label{eq:SCA}
\begin{align}
&\mathop {\min }\limits_{{{\boldsymbol{\phi }}_l},{{\bf{\Omega }}_l}} \sum\nolimits_l {{\boldsymbol{\phi }}_l^H{{\bf{Z}}_l}{{\boldsymbol{\phi }}_l} - 2\Re \left\{ {{\boldsymbol{\phi }}_l^H{\bf{z}}_l^ * } \right\}} \label{eq:SCAo} \\
&\;{\rm{s.t.}}\;\;\;\eqref{eq:FePsRCc1}, \eqref{eq:FePsRCc2}, \label{eq:SCAc1}\\
&\;\;{\rm{Tr}}\left( {{{\bf{D }}_{1,l}}} \right) \le 2{\rm{Tr}}\left( {{\boldsymbol{\phi }}_l^{\left( n \right)}{\boldsymbol{\phi }}_l^H} \right) - {\rm{Tr}}\left( {{\boldsymbol{\phi }}_l^{\left( n \right)}{{\left( {{\boldsymbol{\phi}}_l^{\left( n \right)}} \right)}^H}} \right), \label{eq:SCAc2}
\end{align}
\end{subequations}
which can be solved by a convex optimization solver such as CVX \cite{CVX2012}.{\footnote{For the broadcast scenario with the ES method, the Tx employs one precoding matrix ${\bf{W}} \in {\mathbb{C}^{N \times {{N_{d}}}}}$ to convey the same symbol. Then, by defining ${\bf{U}}_l^ \star  = {\left( {{{{\bf{\bar H}}}_l}{\bf{W}}{\bf{W}}^H{\bf{\bar H}}_l^H + \sigma _l^2{\bf{I}}} \right)^{ - 1}}{{{\bf{\bar H}}}_l}{\bf{W}}$, and ${\bf{V}}_l^ \star  = {{{\bf{\bar H}}}_l}{\bf{W}}{\bf{W}}^H{\bf{\bar H}}_l^H + \sigma _l^2{\bf{I}}$, ${{\bf{Z}}_l} = \left( {{w_l}{\bf{H}}_l^H{{\bf{U}}_l}{{\bf{V}}_l}{\bf{U}}_l^H{{\bf{H}}_l}} \right) \odot {\left( {{\bf{FW}}{{\bf{W}}^H}{{\bf{F}}^H}} \right)^T}$, and ${{\bf{z}}_l} = {\rm{diag}}\left( {{w_l}{\bf{FW}}{{\bf{V}}_l}{\bf{U}}_l^H{{\bf{H}}_l}} \right)$, respectively, we can obtain the corresponding method for broadcast communication with the ES scheme.}}

\section{Extension to the MS and TS Schemes}\label{Ext}
In this section, we consider the MS and TS schemes.

\subsection{Optimization for the MS Scheme}
The main difference between the MS and ES schemes is the constraint $\alpha _m^l \in \left\{ {0,1} \right\}$. Thus, we have the following problem
\begin{subequations}\label{eq:MSPsRC}
\begin{align}
&\mathop {\min }\limits_{{{\boldsymbol{\phi }}_l},{{\bf{\Omega }}_l}} \sum\nolimits_l {{\boldsymbol{\phi }}_l^H{{\bf{Z}}_l}{{\boldsymbol{\phi }}_l} - 2\Re \left\{ {{\boldsymbol{\phi}}_l^H{\bf{z}}_l^ * } \right\}} \label{eq:MSPsRCo} \\
&\;\;{\rm{s.t.}}\;\;\; \eqref{eq:SCAc1}, \eqref{eq:SCAc2}, \label{eq:MSPsRCc1} \\
&\;\;\;\;\;\;\;\;\;\;\;\alpha _m^l \in \left\{ {0,1} \right\},\forall m \in {\cal M}, \label{eq:MSPsRCc2}
\end{align}
\end{subequations}
where \eqref{eq:MSPsRCc2} makes \eqref{eq:MSPsRC} a mixed-integer problem, which is non-convex. To handle \eqref{eq:MSPsRCc2}, we introduce slack variable $\chi _m^l$ and it can be verified that \eqref{eq:MSPsRCc2} is equivalent to $\alpha _m^l = \chi _m^l$, and $\alpha _m^l\left( {1 - \chi _m^l} \right) = 0$ \cite{HuaTWC2021}. Then, we penalize these terms which are included in the OF, and obtain the following problem
\begin{subequations}\label{eq:MSFPsRC}
\begin{align}
\begin{split}
&\mathop {\min }\limits_{{{\boldsymbol{\phi }}_l},{{\bf{\Omega }}_l},{\chi _m^l}} \sum\nolimits_l {{\boldsymbol{\phi}}_l^H{{\bf{Z}}_l}{{\boldsymbol{\phi }}_l} - 2\Re \left\{ {{\boldsymbol{\phi }}_l^H{\bf{z}}_l^ * } \right\}}+ \\
&\;\rho \sum\limits_l {\sum\limits_{m = 1}^M {\left( {{{\left| {\alpha _m^l - \chi _m^l} \right|}^2} + {{\left| {\alpha_m^l\left( {1-\chi _m^l} \right)} \right|}^2}} \right)}}\end{split}\label{eq:MSFPsRCo} \\
&\;\;\;{\rm{s.t.}}\;\;\; \eqref{eq:SCAc1},\eqref{eq:SCAc2}, \label{eq:MSFPsRCc1}
\end{align}
\end{subequations}
where $\rho \ge 0$ is the penalty factor.

For given $\left\{ {{{\boldsymbol{\phi }}_l},{{\bf{\Omega }}_l}} \right\}$, the optimal ${\chi _m^l}$ can be obtained by the first-order optimal condition, which is given by $\chi _m^l = {{\left( {\alpha _m^l + {{\left( {\alpha_m^l} \right)}^2}} \right)} \mathord{\left/
 {\vphantom {{\left( {\alpha_m^l + {{\left( {\alpha _m^l} \right)}^2}} \right)} {\left( {1 + {{\left( {\alpha _m^l} \right)}^2}} \right)}}} \right.
 \kern-\nulldelimiterspace} {\left( {1 + {{\left( {\alpha _m^l} \right)}^2}} \right)}}$ \cite{HuaTWC2021}. On the other hand, for given ${\chi _m^l}$, \eqref{eq:MSFPsRC} can be solved by the previously proposed method.

\subsection{Optimization for the TS Scheme}
For the TS scheme, we propose a two-layer optimization algorithm. Specifically, for given $\tau_l$ and applying the WMMSE scheme as in the ES scheme, we have the following subproblems
\begin{subequations}\label{eq:TSWOp}
\begin{align}
&\mathop {\min }\limits_{{{\bf{W}}_l}} \;\;\sum\nolimits_l {{\tau _l}{\rm{Tr}}\left( {{\bf{W}}_l^H{{\bf{A}}_l}{{\bf{W}}_l}} \right) - 2{\tau _l}\Re \left\{ {{\rm{Tr}}\left( {{{\bf{B}}_l}{{\bf{W}}_l}} \right)} \right\}}  \label{eq:TSWc1} \\
&\;\;\;{\rm{s.t.}}\;\;\eqref{eq:TsOpc1},  \label{eq:TSWc2}
\end{align}
\end{subequations}
and
\begin{subequations}\label{eq:TSPsRC}
\begin{align}
&\mathop {\min }\limits_{{{\boldsymbol{\phi }}_l}} \sum\nolimits_l {{\boldsymbol{\phi  }}_l^H{{\bf{Z}}_l}{{\boldsymbol{\phi }}_l} - 2\Re \left\{ {{\boldsymbol{\phi }}_l^H{\bf{z}}_l^*} \right\}} \label{eq:TSPsRCo} \\
&\;\;\;{\rm{s.t.}}\;\;{\left| {{{\boldsymbol{\phi }}_l}} \right|_m} = 1, \label{eq:TSPsRCoc1}
\end{align}
\end{subequations}
where
\begin{subequations}\label{eq:ABZz}
\begin{align}
&{{\bf{A}}_l} = {w_l}{\bf{\bar H}}_l^H{{\bf{U}}_l}{{\bf{V}}_l}{\bf{U}}_l^H{{{\bf{\bar H}}}_l}, \\
&{{\bf{B}}_l}= {w_l}{{\bf{V}}_l}{\bf{U}}_l^H{{\bf{\bar H}}_l} ,\\
&{{\bf{Z}}_l} = \left( {{w_l}{\bf{H}}_l^H{{\bf{U}}_l}{{\bf{V}}_l}{\bf{U}}_l^H{{\bf{H}}_l}} \right) \odot {\left( {{\bf{F\Xi }}{{\bf{F}}^H}} \right)^T}, \\
&{{\bf{z}}_l} = {\rm{diag}}\left( {{w_l}{\bf{F}}{{\bf{W}}_l}{{\bf{V}}_l}{\bf{U}}_l^H{{\bf{H}}_l}}\right),
\end{align}
\end{subequations}
and ${\bf{U}}_l^ \star  = {\left( {{{{\bf{\bar H}}}_l}{{\bf{W}}_l}{\bf{W}}_l^H{\bf{\bar H}}_l^H + \sigma _l^2{\bf{I}}} \right)^{ - 1}}{{{\bf{\bar H}}}_l}{{\bf{W}}_l}$, ${\bf{V}}_l^ \star  = {{{\bf{\bar H}}}_l}{{\bf{W}}_l}{\bf{W}}_l^H{\bf{\bar H}}_l^H + \sigma _l^2{\bf{I}}$, respectively.

Given $\tau_l$ and ${\boldsymbol{\phi }}_l$, \eqref{eq:TSWOp} can be solved by the Lagrange dual method as shown in \eqref{eq:lag}. On the other hand, given $\tau_l$ and ${\bf{W}}_l$, \eqref{eq:TSPsRC} can be divided into individual subproblems w.r.t ${\boldsymbol{\phi }}_r$ and ${\boldsymbol{\phi  }}_t$. Moreover, each subproblem can be solved by the MM method, which has been widely studied in related works. Specifically, the update of ${\boldsymbol{\phi }}_l$ is given by ${\boldsymbol{\phi }}_l^{\left( {q + 1} \right)} = {e^{j\angle \left( {\left( {{\lambda _l}{\bf{I}} - {{\bf{Z}}_l}} \right){\boldsymbol{\phi }}_l^{\left( q \right)} + {\bf{z}}_l^*} \right)}}$, where $q$ denotes the iteration number, and ${{\lambda _l}}$ denotes the maximum eigenvalue of ${\bf{Z}}_l$ \cite{ZhangTVT20201}.

After obtaining the optimal $\left\{ {{{\bf{W}}_l},{{\boldsymbol{\phi  }}_l}} \right\}$ for given $\tau_l$, the one-dimensional search method is used to obtain the optimal $\tau_l^\star$ and $\left\{ {{{\bf{W}}_l^\star},{{\boldsymbol{\phi }}_l^\star}} \right\}$.{\footnote{For the broadcast scenario with the TS scheme, the method is as the same as the unicast scenario, since whether unicast or broadcast, the TS scheme serves only one user in each time slot.}

\section{Simulation Results}\label{SiRe}
Fig. \ref{Fig:dep} illustrates the considered simulation scenario, where Tx and the STAR-RIS are deployed at $\left( {0,\;0,\;10} \right)$ and $\left( {0,\;30,\;10} \right)$ meters, respectively. The users are randomly deployed on half-circles centered at the RIS with the radius of $5\;{\rm{m}}$, and heights of $2\;{\rm{m}}$. The simulation parameters are given as follows: $N = 4$, ${P_s} = 30~{\rm{dBm}}$, $M=30$, $N_l= 4$, $\sigma _l^2 = -80~{\rm{dBm}}$, ${w_l} = 0.5,\forall l$. The path loss is given by ${\rm{PL}} = {\rm{P}}{{\rm{L}}_0}{\left( {{d \mathord{\left/
 {\vphantom {d {{d_0}}}} \right.
 \kern-\nulldelimiterspace} {{d_0}}}} \right)^{ - \beta }}$, where ${\rm{P}}{{\rm{L}}_0}=10^{-3}$ is the channel gain at ${d_0} = 1\;{\mathrm{m}}$, $d$ is the path distance, and $\beta$ is the path loss exponent. Here, we set $\beta= 2.2$ for all the RIS-related links, similar to \cite{Mu2021}. While the small scale fading channels between Tx and the RIS are modeled as ${\bf{F}} = \sqrt {\frac{{{K_{TR}}}}{{{K_{TR}} + 1}}} {{\bf{F}}^{{\rm{LoS}}}} + \sqrt {\frac{1}{{{K_{TR}} + 1}}} {{\bf{F}}^{{\rm{NLoS}}}}$, where $K_{TR}$ is the Rician factor set as $5\;{\rm{dB}}$. In addition, ${{\bf{F}}^{{\rm{LoS}}}} = {\bf{a}}{{\bf{b}}^H}$ denotes the line-of-sight (LoS) component, where ${\bf{a}} = {\left[ {{a_1}, \ldots ,{a_{N}}} \right]^T}$ and ${\bf{b}} = {\left[ {{b_1}, \ldots ,{b_{M}}} \right]^T}$ denote the transmit and receive steering vectors, respectively. While ${\bf{F}}^{{\rm{NLoS}}}$ denotes the non-LoS component and follows the Rayleigh fading model. The other channels are modeled similarly.
 \begin{figure}[!htb]
\begin{center}
  \includegraphics[width=2.2in,angle=0]{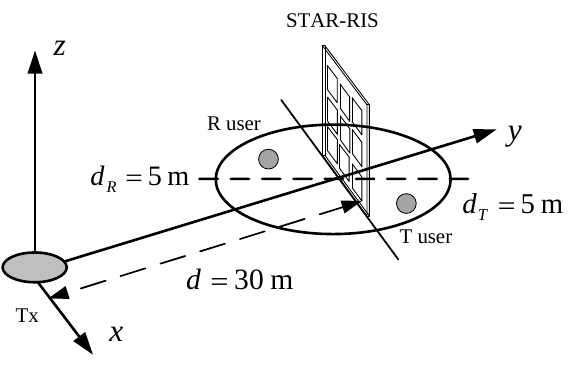}\\
  \caption{The investigated scenario of the STAR-RIS-aided network.}
  \label{Fig:dep}
\end{center}
\end{figure}

Here, we compare our proposed approach with the conventional reflecting-only RIS, where all the incident signal is reflected to the R user and the T user can not receive any information. These methods are labelled as ``ES scheme", ``MS scheme", ``TS scheme", and ``Reflecting-only", respectively.

Firstly, we show the WSR versus the transmit power budget $P_s$ in Fig. \ref{Fig:Ps1}, in the unicast communication scenario. From this figure, we can see that the WSR increases with $P_s$ for all methods, while the TS scheme achieves higher WSR than others, especially in the high $P_s$ region, because it can eliminate the inter-user interference. In addition, the ES scheme achieves higher WSR than the MS scheme, since MS is a special case of ES.
\begin{figure}[!htb]
\begin{center}
  \includegraphics[width=2.5in,angle=0]{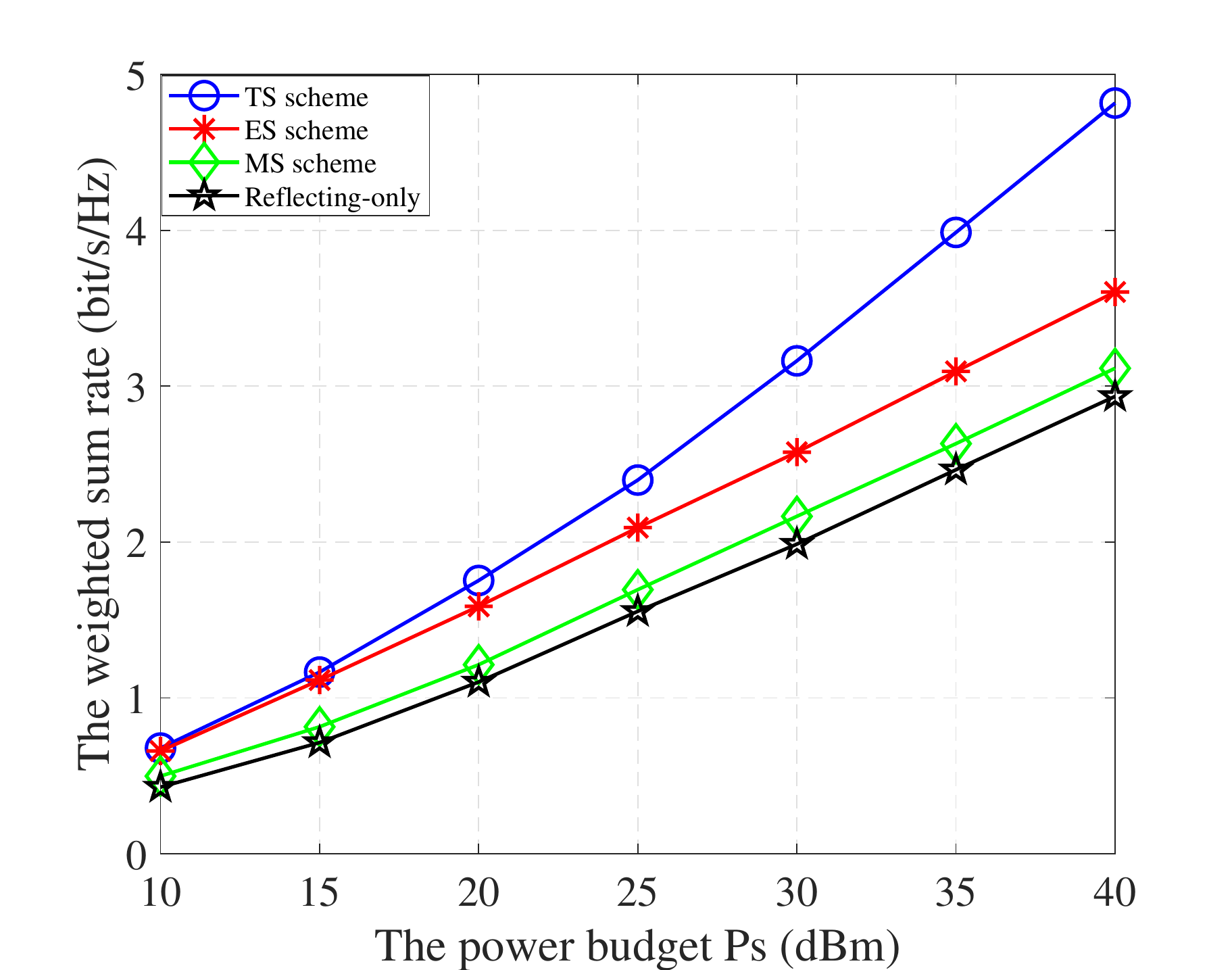}\\
  \caption{The WSR versus $P_s$ in unicast communication.}
  \label{Fig:Ps1}
\end{center}
\end{figure}

Next, we show the WSR versus the number of RIS element $M$ in Fig. \ref{Fig:M}, in the unicast communication scenario. From this figure, we can see that the WSR increases with $M$ for all methods, since with larger $M$, more signals can reach the RIS, and the sum of the transmission and reflection signals increases, provided that the TARCs are properly optimized.
\begin{figure}[!htb]
\begin{center}
  \includegraphics[width=2.5in,angle=0]{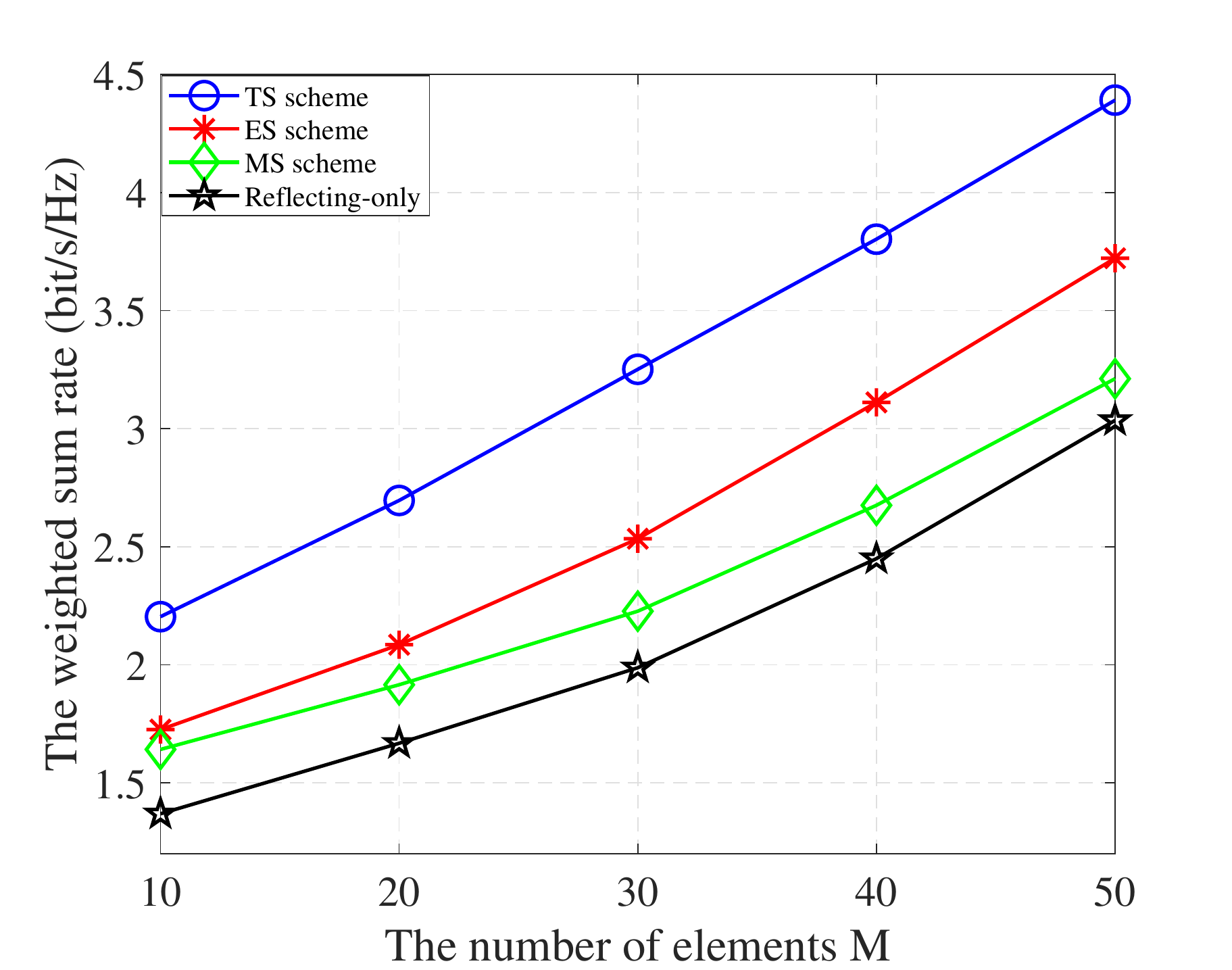}\\
  \caption{The WSR versus $M$ in unicast communication.}
  \label{Fig:M}
\end{center}
\end{figure}

Lastly, we show the WSR versus $P_s$ in Fig. \ref{Fig:Ps2}, in the broadcast communication scenario. Different from the results in Fig. \ref{Fig:Ps1}, in this case, the ES and MS schemes achieve better performance than the TS scheme. This is mainly due to the fact that in the broadcast communication scenario, there is no inter-user interference for the ES or MS scheme, and the two schemes can fully utilize the total available communication time to improve the WSR.
\begin{figure}[!htb]
\begin{center}
  \includegraphics[width=2.5in,angle=0]{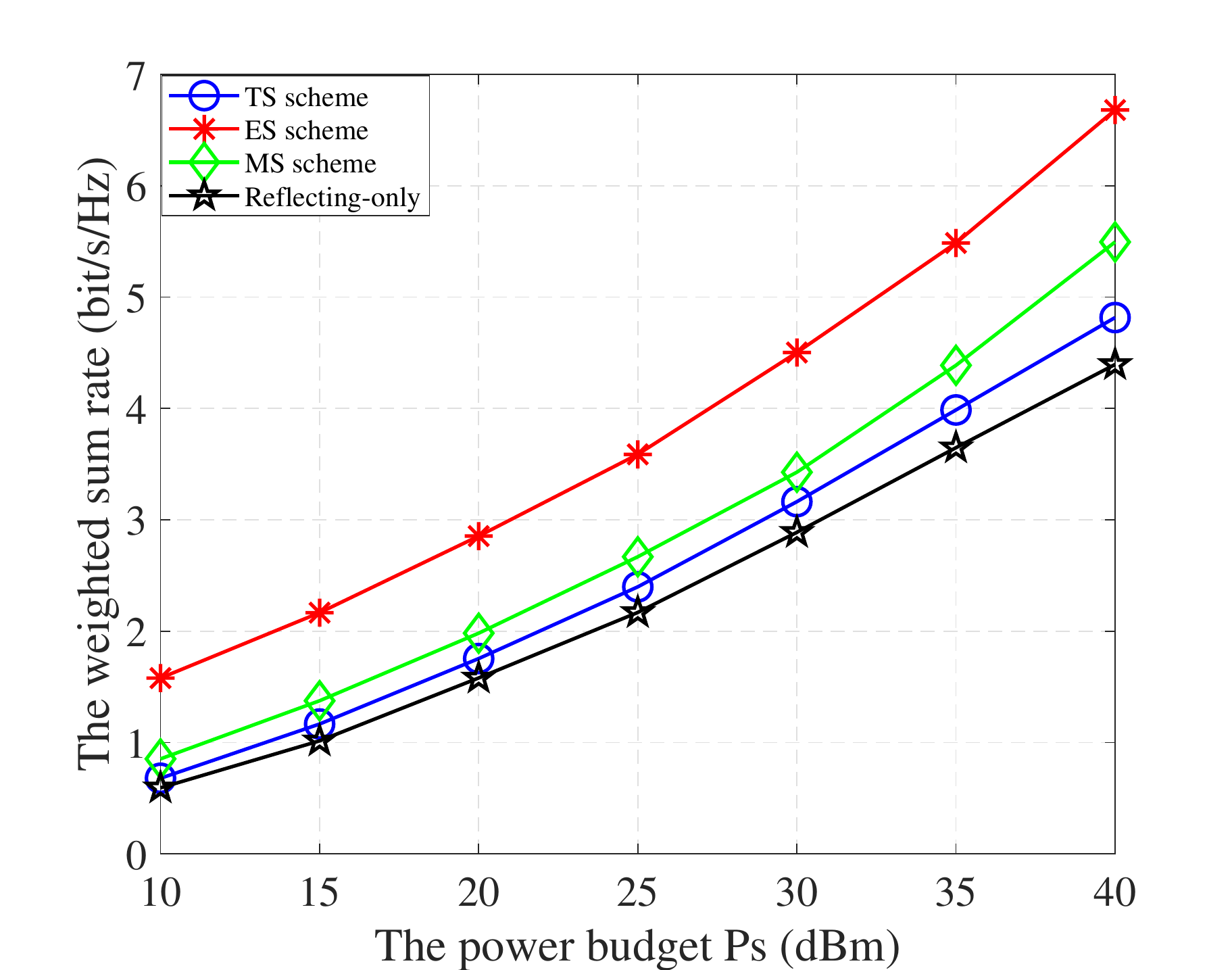}\\
  \caption{The WSR versus $P_s$ in broadcast communication.}
  \label{Fig:Ps2}
\end{center}
\end{figure}

\section{Conclusion}\label{Con}
This work investigated the synergy between the T and R modes for the STAR-RIS-aided MIMO downlink channel. Three different schemes were studied with different problem formulations and corresponding optimization algorithms. Simulation results verified the superiority of the STAR-RIS technique. For our future work, we aim to analyze STAR-RIS-aided designs with hardware impairments. Also, robust design for the STAR-RIS-assisted communication is another direction for future work.

\ifCLASSOPTIONcaptionsoff
  \newpage
\fi

\bibliographystyle{IEEEtran}
\bibliography{IEEEabrv,mybib}

\end{document}